\documentclass[journal]{IEEEtran}

\pdfoutput=1

\usepackage{amsmath}
\usepackage{amsfonts}
\usepackage{bm}
\usepackage{graphicx}
\usepackage{algorithmic}
\usepackage{algorithm}
\usepackage{xcolor}
\usepackage{hyperref}

\usepackage[keeplastbox]{flushend}

\usepackage{import}
\usepackage{multirow}
\usepackage{cite}
\usepackage[export]{adjustbox}
\usepackage{xspace}
\usepackage{balance}
\newcommand{\comment}[1]{#1}

\newcommand{\SM}[1]{#1}
\newcommand{\myparstart}[1]{\noindent \textbf{#1}}
\newcommand{\eq}[1]{Eq.~(\ref{#1})}
\newcommand{\fig}[1]{Fig.~\ref{#1}}

\newcommand{\tab}[1]{Tab.~\ref{#1}}
\newcommand{\secref}[1]{Section~\ref{#1}}
\newcommand{\CodeAddress}{https://github.com/Soldelli/gait\_anomaly\_detection}

\title{Seq2Seq RNN based Gait Anomaly Detection from Smartphone Acquired Multimodal Motion Data}

\author{
    \IEEEauthorblockN{Riccardo Bonetto\IEEEauthorrefmark{1}, Mattia Soldan\IEEEauthorrefmark{4}, Alberto Lanaro\IEEEauthorrefmark{3}, Simone Milani\IEEEauthorrefmark{2}, Michele Rossi\IEEEauthorrefmark{2}\\}
    \IEEEauthorblockA{\IEEEauthorrefmark{1}Institute of Communication Technology, Technische Universit\"{a}t Dresden, 01062, Dresden, Germany\\
    \{riccardo.bonetto\}@gmail.com}\\
    \IEEEauthorblockA{\IEEEauthorrefmark{4}IVUL Lab, King Abdullah University of Science and Technology (KAUST), Thuwal 23955, Saudi Arabia\\
    \{mattia.soldan\}@kaust.edu.sa}\\
    \IEEEauthorblockA{\IEEEauthorrefmark{2}Department of Information Engineering, University of Padova, 35131, Padova, Italy
    \{mattia.soldan.ms\}@gmail.com, \{simone.milani, michele.rossi\}@dei.unipd.it}\\
    \IEEEauthorblockA{\IEEEauthorrefmark{3}M31, 35131, Padova, Italy\\
    \{alb.lanaro\}@gmail.com}}

\begin{document}
\maketitle

\algsetup{
	linenosize=\small,
	linenodelimiter=.
}
\begin{abstract} 
Smartphones and wearable devices are fast growing technologies that, in conjunction with advances in wireless sensor hardware, are enabling ubiquitous sensing applications. Wearables are suitable for indoor and outdoor scenarios, can be placed on many parts of the human body and can integrate a large number of sensors capable of gathering physiological and behavioral biometric information. Here, we are concerned with gait analysis systems that extract meaningful information from a user's movements to identify anomalies and changes in their walking style. The solution that is put forward is {\it subject-specific}, as the designed feature extraction and classification tools are trained on the subject under observation. A smartphone mounted on an \mbox{ad-hoc} made chest support is utilized to gather {\it inertial data} and {\it video signals} from its \mbox{built-in} sensors and \mbox{rear-facing} camera. The collected video and inertial data are preprocessed, combined and then classified by means of a Recurrent Neural Network (RNN) based \mbox{Sequence-to-Sequence} (Seq2Seq) model, which is used as a feature extractor, and a following Convolutional Neural Network (CNN) classifier. This architecture provides excellent results, being able to correctly assess anomalies in $100$\% of the cases, for the considered tests, surpassing the performance of support vector machine classifiers.
\end{abstract}

\section{Introduction}\label{sec:intro}

The automated evaluation of human motion has proven to be a key functionality in different fields, such as \mbox{ambient-assisted} living, remote health monitoring and rehabilitation \cite{Schu17:gait_analysis}, biometric identification~\cite{Gadaleta-2018}, \mbox{well-being} and fitness applications~\cite{har_cnn,har_gmm,har_cvx}. A common trait of these applications is that the effectiveness of Human Activity Recognition (HAR) strategies greatly depends on the continuous and seamless diarization of people's motion and daily activities in different conditions. 
This can be obtained through automatic and \mbox{non-invasive} measuring tools~\cite{sensors}, which are capable of monitoring and recording the motion activity of people with a minimum level of discomfort. 

As a target application, in the present work we consider wearable sensor technology to acquire movement data from heterogeneous sources. Such acquisition setups involve the gathering of a large amount of motion signals, which are to be processed, refined and analyzed to infer some high level information about the condition, the motor status or the identity of a person. Towards this, it is necessary to design flexible and accurate classification tools that are able to include new subjects in the analysis (\emph{open set}) and characterize their motion accurately. Motion analysis is a vivid research field, and people gait is usually investigated using 3D cameras, which allow for an accurate tracking of the trajectory of the skeleton and the joints. In this work, we depart from the previous literature as our objective is to use \mbox{low-cost} and \mbox{non-specialized} sensing hardware, i.e., a smartphone. The amount of information that we gather is much more limited than with 3D video systems, but nevertheless, we would like to infer some useful information about the motion \comment{behavior of the monitored subject.} Our technology may eventually be useful for the quick assessment of whether a certain motion disorder is emerging and/or for personalized sport applications (e.g., running). 

\comment{
Automatic gait analysis approaches can be divided into three main classes depending on the characteristics of the sensing devices~\cite{Mur14:gait_sensors}. A first class concerns \mbox{vision-based} systems, where subjects are monitored and analyzed by a set of fixed cameras in a controlled environment~\cite{Gof2010:mcam,Lit96:gait}. Although these solutions are extremely accurate in characterizing the motion of human subjects, they lack flexibility since they are heavily affected by changes in illumination and occlusions, and they also need some highly specialized and expensive equipment, \comment{which requires adequate space for its deployment and expert personnel for its operation.} A second class of techniques involves environment interactive sensors~\cite{Gee16:mov_disord,mid05:sensor_floor}, i.e., sensing devices that depend on the specific equipment used by the analyzed subject, e.g., a sport gear, \mbox{infra-red} reflective tags, pressure sensing mats for plantar pressure analysis, etc. \comment{Their use is often appropriate for indoor spaces and for a restricted number of activities~\cite{device_free}. For example, in~\cite{Pogorelc-2010} the gait of a monitored subject is captured using infrared (IR) cameras in conjunction with reflective tags worn in several parts of the body (usually at the joints), detecting the following anomalous patterns: hemiplegia, Parkinson (shuffling walk), leg and back pain.}
A third class of solutions entails the use of wearable motion sensors~\cite{Tao12:wearable}, which can be employed anywhere, anytime, and with limited discomfort for the wearer. In this case, accelerometers, gyroscopes, magnetometers, GPS and other kind of devices are placed on different parts of the body~\cite{smart_insole}; during the walking activity, signals are measured and stored for their subsequent classification~\cite{Mar12:intertial_gait}. \comment{The study in~\cite{Cola-2015} uses a \mbox{Shimmer 2} \mbox{tri-axial} acceleration sensing device worn at the subject's lower back to detect mild and severe knee conditions. Walking patterns are extracted, segmented into subsequent walk cycles, and assessed using a binary \mbox{k-Nearest Neighbors (k-NN)} classifier.} Despite their flexibility, the accuracy of these solutions is limited by high noise levels and by the fact that often the reference system depends on the orientation of the sensing device, which is not fixed and may have to be (re)estimated at measurement time. 

A large body of work has been recently published on multimodal data processing for gait analysis~\cite{els10:gait_ambient_wearable,Hos12:multimodal_gait}, by especially investigating the supervised classification of walking activity. Most of the solutions transform the input data through Principal Component Analysis (PCA)~\cite{Mar12:intertial_gait} and \mbox{noise-removal} strategies~\cite{Soa11:gait_denoise}, moving it into a space representation that is suitable for classification. Then, data is classified using machine learning strategies like \mbox{k-d trees}~\cite{Mar12:intertial_gait}, Support Vector Machines (SVMs)~\cite{Nak16:gait_svm}, and Artificial Neural Networks (ANNs)~\cite{Nuk16:gait_ann}. Recently,  Recurrent Neural Networks (RNNs) and RNN-based sequence to sequence models have become popular for applications such as speech to text~\cite{STT, STT2}, text to speech~\cite{TTS}, sentiment analysis~\cite{RNN-SA1, RNN-SA2}, and neural machine translation~\cite{NMT1,NMT2}. Their increasing adoption is mostly due to the architectural flexibility of neural network models in conjunction with the ability of RNNs to encode into state vectors the temporal information underpinning \mbox{multi-dimensional} timeseries.

To the best of the authors' knowledge, the aforementioned advances in RNN based \mbox{Seq-2-Seq} learning have not yet been applied to the classification of gait signals. In this paper, we aim at filling this gap. The proposed classification strategy identifies anomalous gaits that may happen during a normal walk. The anomalies that are detected can then be the subject of further analysis, \comment{i.e., to further split the anomaly into a number of anomaly classes}, by means of standard techniques~\cite{MDS,MDS2,MDS3}.
}\\

\noindent \textbf{Contributions of the paper.} We propose a \mbox{\it subject-specific} gait anomaly detection framework that combines recent advances in recurrent neural network based sequence to sequence (Seq-2-Seq)~\cite{GoogleMTS_16}\cite{Seq2Seq_Speech_18} models with the largely proven classification abilities of Convolutional Neural Networks (CNN). We adopt a \mbox{multi-modal} gait analysis approach, which integrates inertial and visual data. Motion is extracted using {\it accelerometric and gyroscopic measurements} from a smartphone device that moves integrally with the analyzed subject. These data are merged with the {\it optical flow information} obtained from an ego vision system corresponding to the smartphone \mbox{built-in} camera. Then, the gathered data traces are automatically segmented into gait cycles, which undergo a filtering, a detrending, and a normalization process. The refined multi-variate timeseries are then embedded into a feature space obtained through a \mbox{sequence-to-sequence} architecture based on a recurrent neural network (herein referred to as \mbox{RNN-Seq2Seq}). These embeddings are then fed to a CNN binary classifier.

The RNN-Seq2Seq model is set to echoing the input sequence like an autoencoder. Once a sequence has been fully processed, the final state of the RNN encoder is extracted and reshaped into a multidimensional matrix. This, in turn, is fed to a CNN based classifier that has the task of separating ``normal'' gaits from ``anomalous'' ones. In this work, anomalies are defined with respect to the normal walking style of the subject under analysis. For this reason, preexisting and known conditions affecting the walking pattern of the analyzed subject (as, for example, old knee or hip injuries) are not treated as anomalies that have to be detected. The \mbox{RNN-Seq2Seq} is only trained on ``normal'' (i.e., usual) gaits with the purpose of making the encoder unable to correctly embed ``anomalous'' gaits in its feature space. Hence, the difference between the features of ``normal'' and ``anomalous'' gaits are expected to be amplified, thus, facilitating the classification task. 

The output of the trained recurrent model is subsequently inputted into the CNN classifier, which is trained on a different dataset containing \mbox{pre-labeled}  ``normal'' and ``anomalous'' gaits (with equal cardinality). Once the training procedure for the RNN and the CNN is concluded, the full system is tested for classification accuracy on a validation dataset. As a means of comparison, we also implemented a \mbox{non-linear} Support Vector Machine (SVM) model, which has been trained on the same training set used for the CNN classifier, and has been fed with complete gait cycles.

Experimental results show that  the proposed RNN-Seq2Seq pre-encoding and the subsequent CNN-based classification outperform the baseline accuracy obtained by SVM, achieving an accuracy of $100$\% in the detection of anomalous gaits in our own collected dataset. \comment{The source code of our framework for gait analysis, along with preprocessed data and the signal acquisition system from Android smartphones is publicly released and available at: \href{\CodeAddress}{\CodeAddress}.} 

The rest of this paper is organized as follows. The signal acquisition strategy, and the data \mbox{pre-processing} operations are presented in \secref{sec:dp}. The proposed \mbox{RNN-Seq2Seq} model is discussed in \secref{sec:emb}. The proposed CNN based classifier is detailed in \secref{sec:class}. The dataset creation, its characteristics, along with implementation details and parameters of all the considered approaches are presented in \secref{sec:es}. The experimental results are shown in \secref{sec:res} and some final considerations are drawn in \secref{sec:conc}.

\section{Materials and Methods}\label{sec:mm}
\comment{Here, we introduce our anomaly detection approach and define the experimental setup used for its validation.}

\subsection{Data Processing}\label{sec:dp}

\begin{figure}[t]
\hspace{1.5mm}
\centerline{\begin{minipage}[c]{0.41\columnwidth}
\includegraphics[width=\columnwidth]{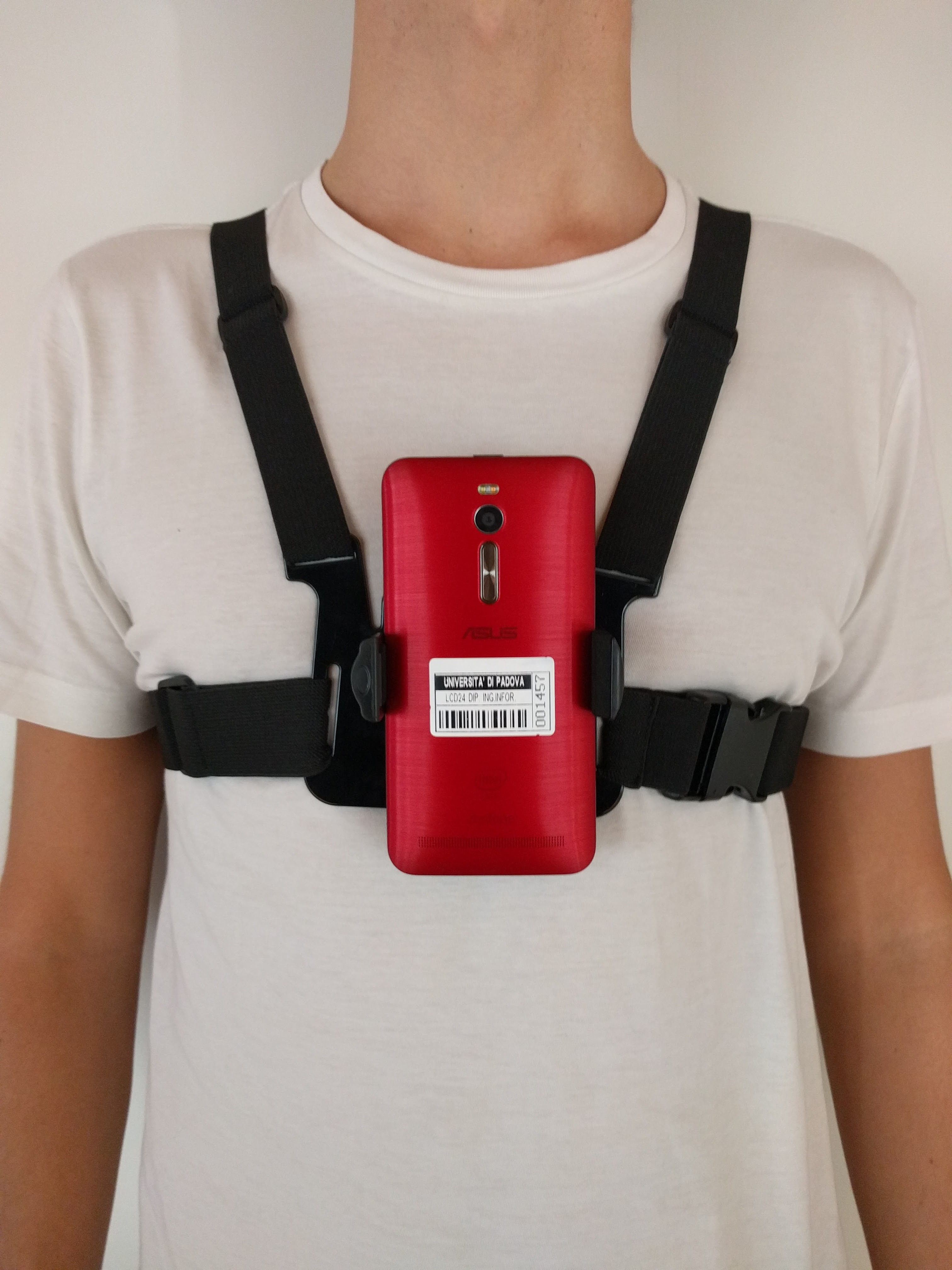} \\
\centering{(a)}
\end{minipage} \hfil
\begin{minipage}[c]{0.55\columnwidth}
\includegraphics[width=\columnwidth]{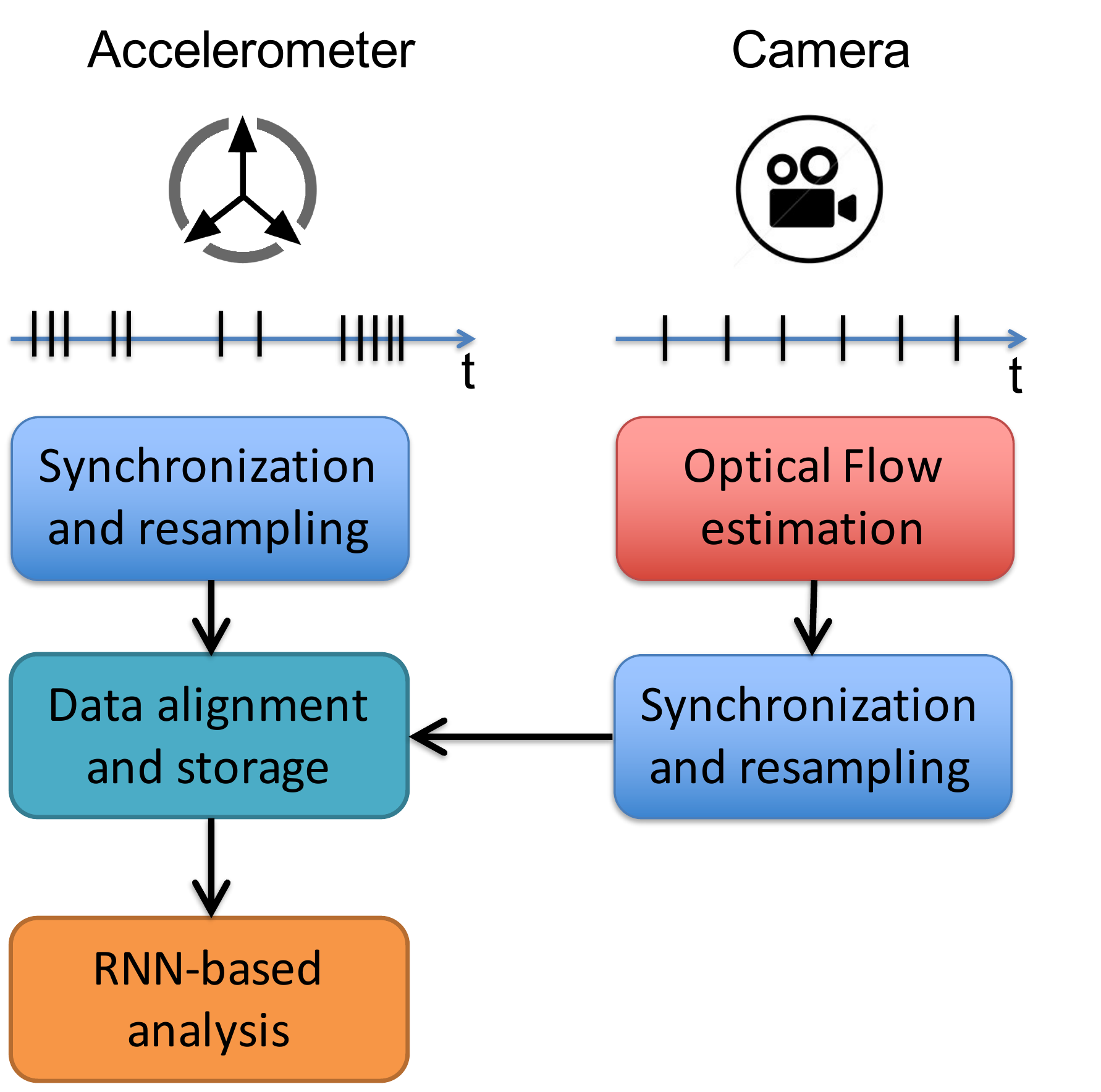}  \\
\centering{(b)}
\end{minipage}}
\caption{(a) smartphone chest support; (b) functional block diagram.}
\label{fig:chest_support}
\end{figure}

As with previous work on automatic gait analysis, the proposed system requires the acquisition of accurate motion data for the subject that is to be monitored. To gather this information, we adopted a \mbox{multi-modal} approach that combines {\it video} and {\it inertial signals}. As a relatively cheap and convenient way to acquire this data, we opted for a smartphone application, since camera and inertial sensors are already integrated into the device. Hence, we developed a chest support for smartphones (showed in Fig.~\ref{fig:chest_support}(a)) and a motion sensing application \SM{that} records accelerometer, gyroscope, and magnetometer measurements, while also recording a video sequence from the front camera. A block diagram of the smartphone data acquisition and processing system is shown in Fig.~\ref{fig:chest_support}(b). Note that, prior to signal classification and clustering, the data coming from different sensors is aligned, denoised, and processed to extract salient motion information. In the following, the processing blocks are described in detail.\\

\noindent \textbf{1) Inertial Data Acquisition and Synchronization:} Inertial signals are provided by built-in gyroscopic, magnetometric and accelerometric sensors. At each sampling epoch, each of these sensors returns a \mbox{three-dimensional} real sample ($3$-axes, for a total of $9$-axes for the three sensors), related to the motion of the device along the three dimensions of the smartphone reference system. Inertial samples are labelled by a timestamp that is relative to the smartphone's system clock. Although the sampling frequency is high (typically around $100/200$~samples/s, depending on the device), the time interval between consecutive samples is not constant. \comment{This is consistent with the findings of \cite{frequency_stability}}. To cope with this, interpolation and resampling are performed prior to data analysis to convert the signals into the common sampling frequency of $200$~Hz, \comment{as done in prior work~\cite{200hz}}.
Moreover, the power spectral density of the accelerometer signals show that sensor samples are affected by a significant amount of noise due to the irregularities of motion and to the sensitivity of the sensing platform. This noise is largely removed through a low pass filter with a \mbox{cut-off} frequency of $40$~Hz.\\

\noindent \textbf{2) Video Data Extraction:} Besides inertial measurements, motion data is also extracted from a video sequence that is concurrently acquired during each walk. At time instant $n$, the phone/camera pose can be represented by a location vector $\bm{t}_n$ and an orientation matrix $\bm{R}_n$, which can be incrementally obtained through the relative rotation $\hat{\bm{R}}_n$ and translation $\hat{\bm{t}}_n$, i.e., $\bm{R}_n = \hat{\bm{R}}_n \bm{R}_{n-1}$ and $\bm{t}_n = \hat{\bm{R}}_n \bm{t}_{n-1} + \hat{\bm{t}}_{n}$.
As a result, the same points in any two adjacent frames $I_n$ and $I_{n-1}$ would appear modified by an affine transformation characterized by $\hat{\bm{R}}_n$ and $\hat{\bm{t}}_n$, which can be estimated as follows.
\setlength{\unitlength}{1mm}
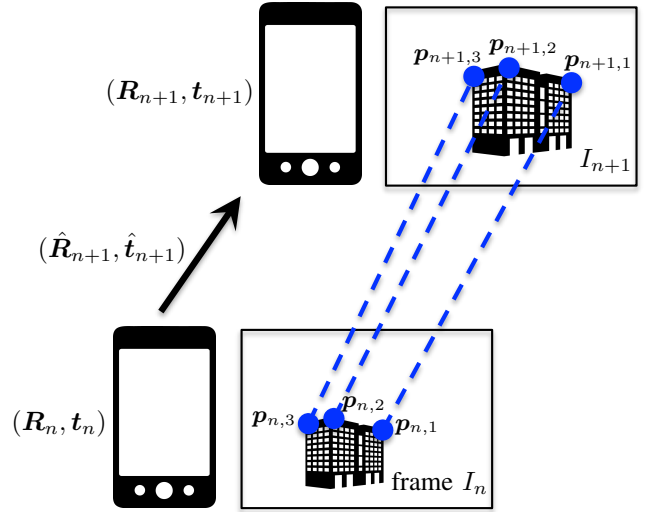
\begin{figure}[t]
\begin{center}
\begin{picture}(85,70)(0,0)
\put(15,0){\includegraphics[width=0.8\columnwidth]{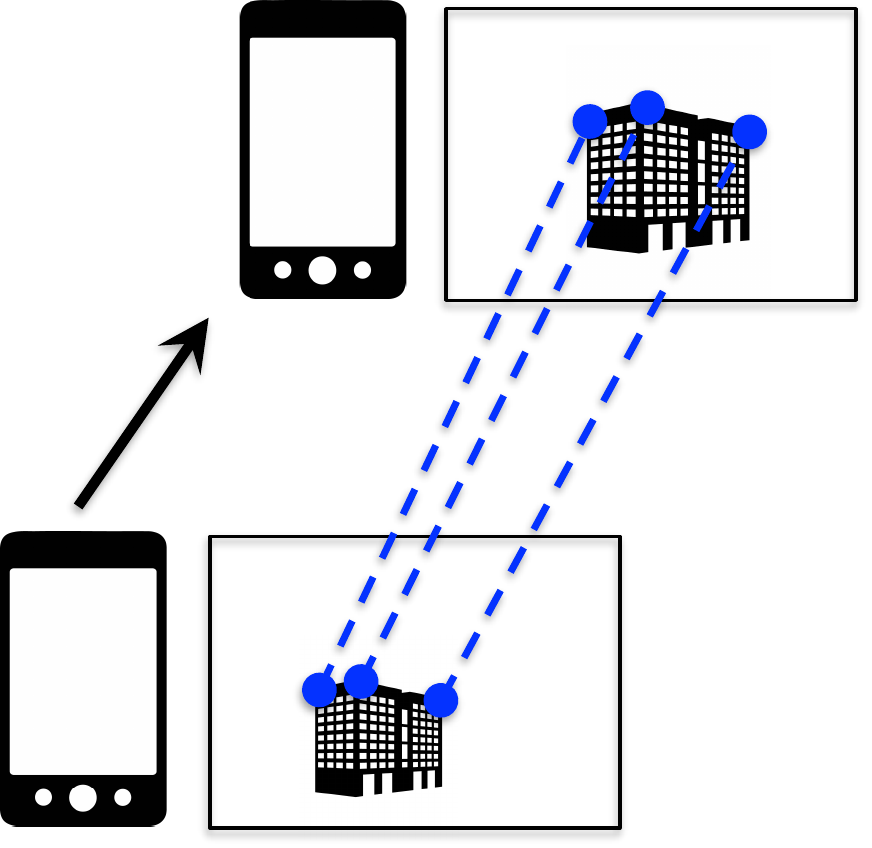}}
\put(1.5,12){$(\bm{R}_{n}, \bm{t}_{n})$}
\put(5,35){$(\hat{\bm{R}}_{n+1}, \hat{\bm{t}}_{n+1})$}
\put(14,56){$(\bm{R}_{n+1}, \bm{t}_{n+1})$}
\put(52,4){frame $I_n$}
\put(77,47){$I_{n+1}$}
\put(75,60.4){\small $\bm{p}_{n+1,1}$}
\put(65,62.5){\small $\bm{p}_{n+1,2}$}
\put(55,61.2){\small $\bm{p}_{n+1,3}$}
\put(52.5,12){\small $\bm{p}_{n,1}$}
\put(45.4,15){\small $\bm{p}_{n,2}$}
\put(33.5,13){\small $\bm{p}_{n,3}$}
\end{picture}
\end{center}
\caption{Relation between corresponding points in adjacent video frames.}
\label{fig:rel_video}
\end{figure}
For every video frame $I_n$, acquired at time instant $n$, the video processing unit computes a set \SM{$\mathcal{D}_n$} of keypoints \SM{that can be easily tracked across subsequent frames} (see Fig.~\ref{fig:rel_video} for a graphical example), i.e.,
\begin{equation}
\mathcal{D}_n=\{ \bm{p}_{n,k}=(x_{n,k},y_{n,k},1) \} \,  , 
\end{equation}
\noindent \SM{where $\bm{p}_{n,k}$ are expressed in homogeneous normalized coordinates}. \SM{Salient points were identified using the SIFT algorithm~\cite{Lowe04:sift},  which allows one to select a set of pixel patches that are \mbox{scale-invariant} across frames}. \SM{The Farneback optical flow algorithm~\cite{Farn03:optical_flow} is then applied on frames $I_n$ and $I_{n-1}$; as a result, the  point $\bm{p}_{n,k}$ of  frame $I_n$  is associated with a point $\bm{p}_{n-1,k}$ of the previous frame $I_{n-1}$. }

\SM{These correspondences allow the estimation of the $3 \times 3$ {\it essential matrix} $\bm{E}_n$, which satisfies the Longuet-Higgins equation,} 
\begin{equation}
0 = \bm{p}_{t,k}^T \ \bm{E}_n \ \bm{p}_{n-1,k} =\bm{p}_{n,k}^T \ [\hat{\bm{t}}_n]_{\times} \hat{\bm{R}}_n \ \bm{p}_{n-1,k} . 
\end{equation}
\SM{In this case, the matrix  $\bm{E}_n$ is factored into the product of the vector product matrix $ [\hat{\bm{t}}_n]_{\times} $ (associated with the relative translation vector $\hat{\bm{t}}_n$) and the relative rotation matrix $\hat{\bm{R}}_n$. Following this, $\bm{E}_n$ is estimated through the Nister's \mbox{$5$-point} algorithm combined with a RANSAC optimization procedure to remove false matches and outliers. The matrix $\bm{E}_n$ can be decomposed into $\hat{\bm{t}}_n$ and $\hat{\bm{R}}_n$ via an \mbox{SVD-based} factorization. \comment{This makes it possible to estimate} the location and the orientation of the smartphone camera at instant $n$, composing them with $\bm{R}_{n-1}$ and $\bm{t}_{n-1} $.}
From the resulting rotation matrix $\hat{\bm{R}}_n=\left[r_{i,j}\right]$, we obtain the roll, pitch, and yaw angles of the camera, $\alpha_n$, $\beta_n$ and $\gamma_n$, which are defined with respect to the reference system associated with the camera pose at instant $n$. These angles are obtained from the rotation matrix $\hat{\bm{R}}_n$ as:
\begin{equation}
\label{eq:rpy}
\begin{array}{l}
\displaystyle \alpha_n=\tan^{-1}\left( \frac{r_{2,1}}{r_{1,1}} \right) \, , \, \gamma_n = \tan^{-1} \left( \frac{r_{3,2}}{r_{3,3}} \right) \, , \\\displaystyle \beta_n = \tan^{-1} \left( \frac{-r_{3,1}}{\sqrt{(r_{3,2})^2 + (r_{3,3})^2}}\right) \, .
\end{array}
\end{equation}
The estimation accuracy depends on the capture frequency of video frames: for our experiments we used a frame rate of $30$~Hz, \comment{which is consistent with that adopted in previous studies, e.g.,~\cite{Bouchrika-2011}.}  At the beginning of the video acquisition, a timestamp $\tau_0$ is recorded (whose value depends on the smartphone's system clock). Since the sampling frequency is constant, a rather precise acquisition time, computed as $\tau_0 + n \Delta$, can be associated with every acquired frame, i.e., with every estimated triplet of angles $(\alpha_n, \beta_n, \gamma_n)$, where $\Delta$ is the sampling period.

At this point, it is necessary to synchronize the estimated angles with the signals acquired by the inertial sensors. Since the time resolution of the inertial data is finer, \mbox{video-related} samples are interpolated to a sampling frequency of $200$ Hz, and are time synchronized with the inertial data (estimating the optimal \mbox{delay-shift} that aligns video and inertial traces).\\

\begin{figure}
\center
\includegraphics[width=\columnwidth]{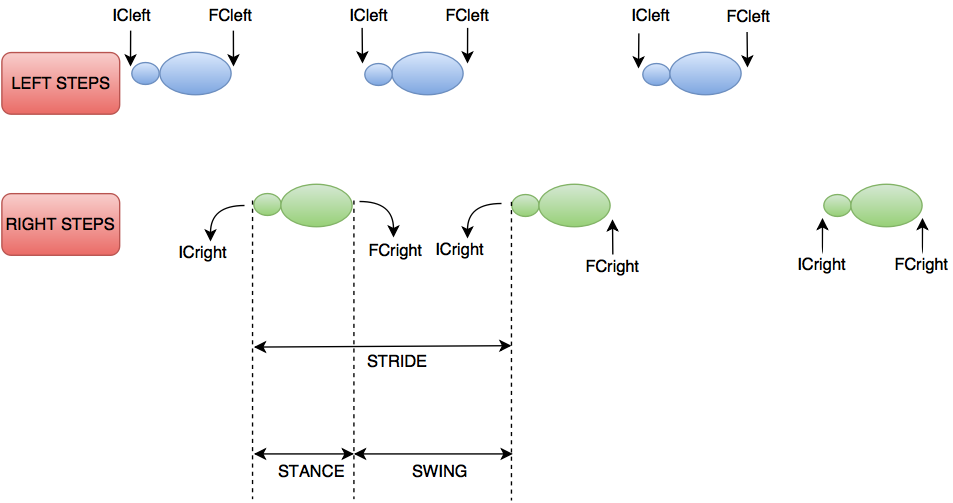}
\caption{Simplified diagram of walking phases, Initial Contact (IC) and Final Contact (FC) instants are marked for right and left steps.}
\label{fig:icfc_diagram}
\end{figure}

\noindent \textbf{3) Gait Cycle Extraction:} The human gait follows a {\it cyclic} behavior featuring a periodic repetition of a pattern delimited by two consecutive steps. The {\it stance} phase starts with the instant when contact is made with the ground (usually occurring with the heel touching the ground first); this instant is called Initial Contact (IC). After that, the foot becomes flat on the ground and supports the full body weight (foot flat). Then, the heel begins to lift off the ground in preparation for the forward propulsion of the body and we finally have the take off phase, which ends the stance and is delimited by the instant of Final Contact (FC) of the foot with the ground. Afterwards, the weight of the body is moved to the other foot until the next IC occurs ({\it swing} time).
\comment{A gait cycle ({also referred to as \it stride or \it {walking cycle}}) is defined as the time instants between two consecutive Initial Contacts (ICs) of the same foot}. A pictorial representation of ICs and FCs is given in \fig{fig:icfc_diagram}. IC and FC instants can be identified by analyzing the vertical component of the accelerometer data. To this end, the signal is processed by a Difference of Gaussians (DoG) filter followed by a wavelet transform. IC instants correspond to the local minima of the transformed signals, while FC ones are found through a second differentiation and searching for local maxima. To avoid false detections, only IC and FC events within specific time intervals are considered~\cite{DelDin2016:gait}.

\begin{figure}
\center
\includegraphics[width=\columnwidth]{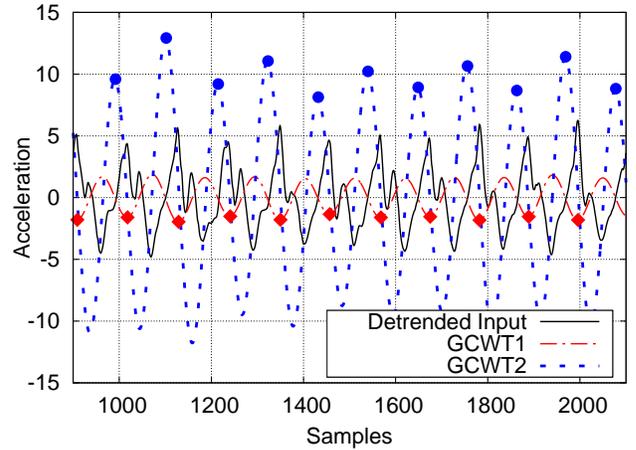}
\caption{Initial (IC) and final (FC) contact instants for an example accelerometer signal. GCWT means Gaussian Continuous Wavelet Transform.}
\label{fig:icfc}
\end{figure}

After this processing phase, gait cycles can be reliably identified. In fact, a generic walking cycle $i$ starts at IC$(i)$ and ends at IC$(i+2)$. It is thus possible to locate the walking cycles vectors in all the available signals. Each {\it gait cycle} is then normalized to a fixed length of $200$ samples, stored on a descriptor and classified with the technique of Section~\ref{sec:emb}. An example of IC and FC detection is shown in \fig{fig:icfc}.\\

\begin{figure}[t]
\includegraphics[width=\columnwidth]{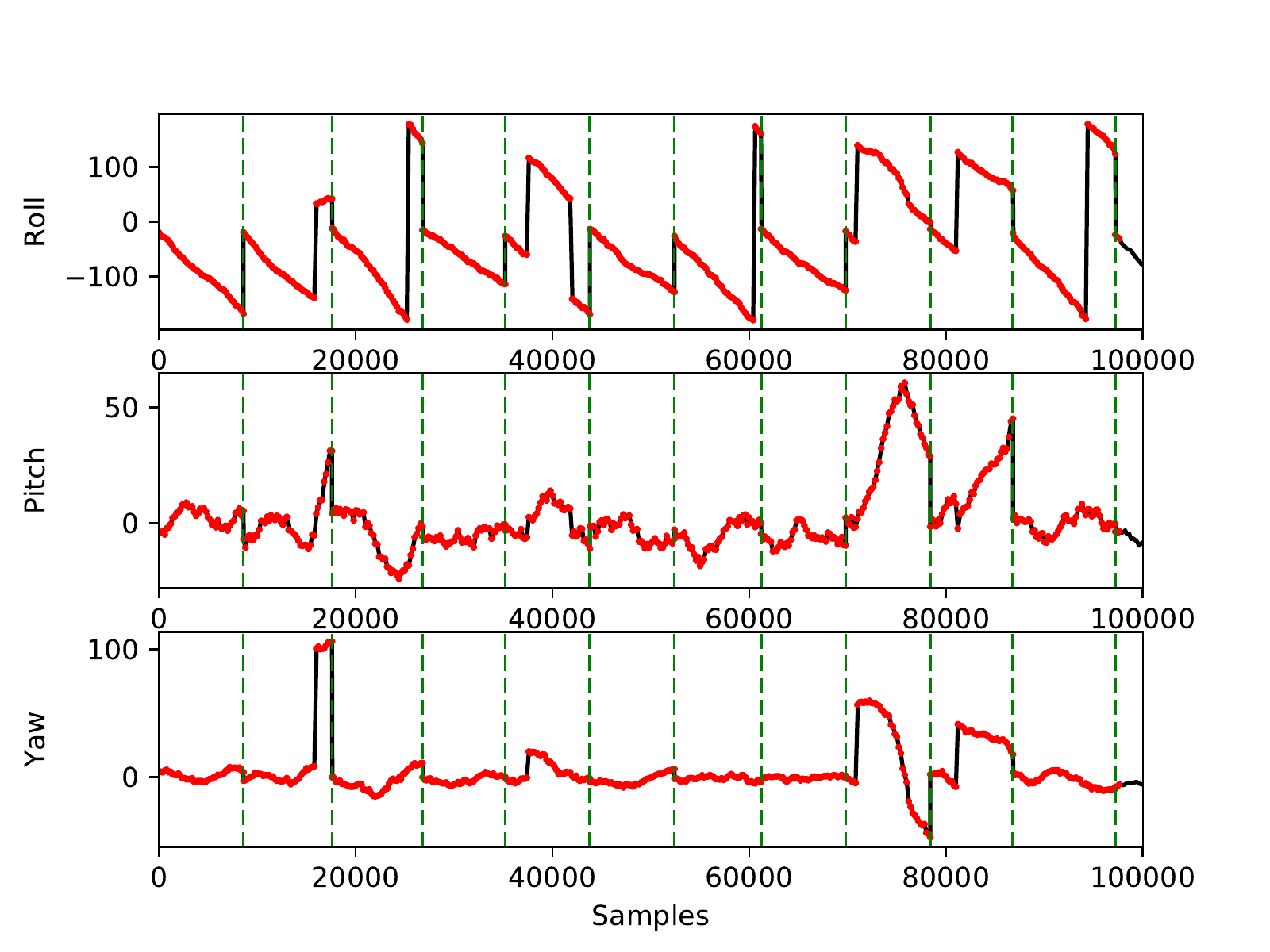}
\caption{Example of trends in the Roll, Pitch, and Yaw signals extracted from the video acquisition. The vertical lines identify the gait segments.}
\label{fig:trend}
\end{figure}

\noindent \textbf{4) Detrending:} \comment{The signals extracted from the reference video have trend components on each of the three axes, which can heavily impact the data normalization, see Fig.~\ref{fig:trend} for examples. To remove such trends and obtain \mbox{semi-stationary} timeseries, we have fitted a linear model for each gait cycle. Then, the slope of the computed model has been used to remove the trend from the corresponding gait cycle}. Therefore, the trend affecting each cycle is approximated by a linear model. This permits to remove the global trend without affecting the gait cycles. 

\begin{figure}[t]
\includegraphics[width=\columnwidth]{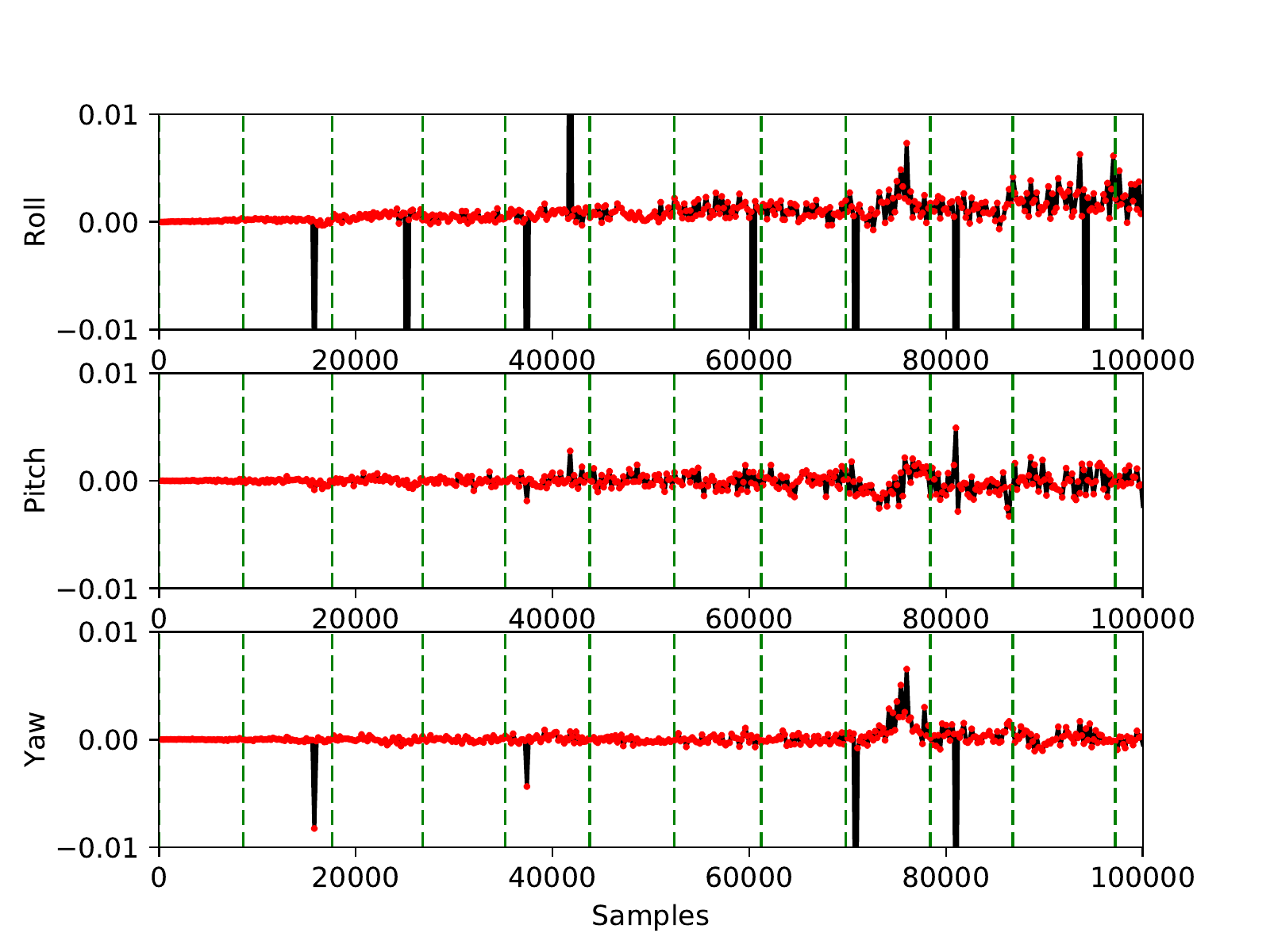}
\caption{Example of the detrended Roll, Pitch, and Yaw signals extracted from the video acquisition. The vertical lines identify the gait segments.}
\label{fig:detrend}
\end{figure}

\comment{The result of the detrending procedure can be seen by comparing \fig{fig:trend}, and \fig{fig:detrend}.} \fig{fig:trend} shows an example of the trends that affect the roll, pitch, and yaw signals extracted from the video. \fig{fig:detrend} shows the result of detrending. \SM{It is worth noting that spurious peaks still remain after this operation, but they can be promptly removed using an additional threshold based peak detection algorithm.}\\


\noindent \textbf{5) Normalization}
Data normalization is a standard step in data preparation for neural network based learning. It is performed by removing the mean and dividing the data by its standard deviation, \comment{computing these measures over the entire dataset.} Here, this operation is performed for each of the $9$ signals in a gait cycle.

\subsection{Gait RNN Based Embedding}\label{sec:emb}

Next, we present an RNN based architecture to embed the gait signals into a feature space containing a fixed size, higher order representation of the gaits.\\

\noindent \textbf{Introduction to Sequence to Sequence Models:} RNN based Sequence to Sequence (RNN-Seq2Seq) models have gained a lot of attention lately, see, e.g.,~\cite{Seq2Seq_Speech_18}, mainly thanks to the work on Neural Machine Translation (NMT) by Google~\cite{GoogleMTS_16}. NMT models are based on a 2 blocks architecture featuring an Encoder (first block) and a Decoder (second). The encoder is an RNN that, when fed with an input sequence of words (i.e., a sentence, possibly of variable lentgh), embeds it into a fixed size feature space. This embedding, being generated by an RNN, captures temporal correlations between different portions of the input sequence. The embedded input sequence is then fed to the decoder RNN that, in turn, generates an output sequence that corresponds to the translation of the input. In this work, we design an RNN-Seq2Seq \mbox{encoder-decoder} architecture to embed \mbox{multi-dimensional} \mbox{fixed-length} timeseries into a feature space that is suitable for a subsequent classification task. To achieve this, we utilize a deep RNN as the encoder and a shallow NN with linear activations as the decoder. 

\begin{figure*}[t]
\center
\def\svgwidth{1.8\columnwidth}
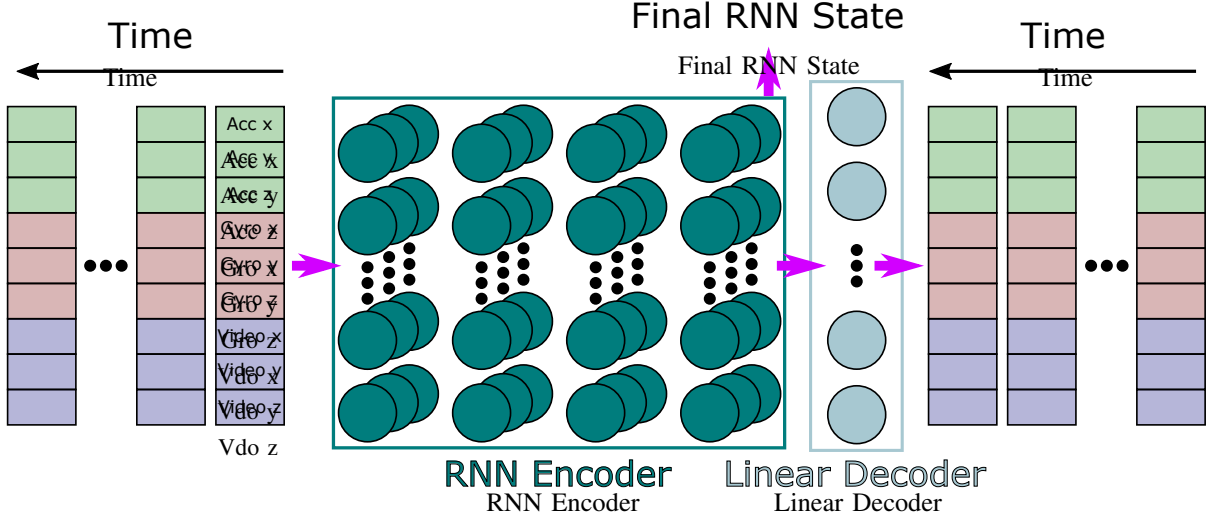
\vspace{0.1cm}
\caption{RNN-based Embedding Architecture. ``Acc'' stands for Accelerometer, ``Gro'' stands for Gyroscope, and ``Vdo'' stands for Video. }\label{fig:rnn_encoder}
\end{figure*}

In \fig{fig:rnn_encoder}, we show the architecture of the designed \mbox{Seq-2-Seq} model. First, the \mbox{multi-dimensional} timeseries corresponding to a gait cycle is fed into the RNN encoder. Then, the output of the encoder is fed to the shallow linear decoder. The whole neural network chain is trained to reproduce the input sequence at its ouput. The final state of the encoder, is then utilized as a feature vector representing the embedding of the input sequence into a {\it feature space} that retains temporal information about the processed gait cycle.\\

\noindent \textbf{Deep RNN Encoder:} To obtain a finite-dimensional representation of the temporal characteristics of the gait multivariate timeseries, we utilized a deep bidirectional Long Short Term Memory (LSTM) based design~\cite{Bidirectional_LSTM_1997,Deep_Birectional_LSTM_2015}. Each recurrent layer produces a higher order encoding (i.e., spanning a longer temporal information) of the output of the previous one. This is due to the fact that each subsequent layer, apart from the first one, receives as input an already \mbox{time-dependent} and encoded representation of the original multivariate timeseries. The bidirectional connections allow updating the past network states according to what will be happening into the future. The number of dimensions of each encoding is determined by the size of the corresponding layer. Hence, the depth of the encoder determines the order of the statistics represented by the features that are extracted from the input sequence, while the size of the last layer determines the size of the space in which the original sequence is finally embedded.

Formally, let $\bm{X} = [\bm{x}_0,\bm{x}_1,\dots,\bm{x}_N]$, $\bm{X} \in \mathbb{R}^{9\times {N+1}}$ be the timeseries representing a gait cycle, where $\bm{x}_i\in\mathbb{R}^9$, \mbox{$i = 0,\dots,N$}, is the preprocessed motion vector at sampling instant $i$, \comment{combining inertial signals (accelerometer and gyroscope) and the motion angles (roll, pitch and yaw) obtained from the video.} Then $\bm{X}$ is fed to the encoder, one sample at a time, until $\bm{x}_N$ is processed. When sample $\bm{x}_i$ is processed, the layer $\ell=0,\dots,L-1$ of the encoder produces a representation $R_\ell(\bm{X}_{0,i})$ of the subsequence \mbox{$\bm{X}_{0,i} = [\bm{x}_0,\dots,\bm{x}_i]$}. Given the recurrent nature of this architecture, the obtained representation vector, at the output of the RNN Encoder, can be expressed as:
\begin{equation}
R_\ell(\bm{X}_{0,i}) = \bm{f}(\dots(\bm{f}(\bm{X}_{0,i},\bm{W}_0,\bm{S}_0,\bm{b}_0),\dots),\bm{W}_\ell,\bm{S}_\ell,\bm{b}_\ell) \,
\label{eq:representation}
\end{equation}
where $\bm{W}_\ell$ is the weight matrix, $\bm{S}_\ell$ is the state matrix, and $\bm{b}_\ell$ is the vector of biases, associated with layer $\ell$. Moreover, let $\bm{f}(\cdot,\bm{W}_\ell, \bm{S}_\ell, \bm{b}_\ell)$ be:
\begin{equation}
\bm{f}(\cdot,\bm{W}_\ell,\bm{S}_\ell,\bm{b}_\ell) = 
\begin{pmatrix}
f(\cdot,\bm{w}^0_\ell,\bm{s}^0_\ell,b^0_\ell) \\
\vdots \\
f(\cdot,\bm{w}^U_\ell,\bm{s}^U_\ell,b^U_\ell) \\
\end{pmatrix}\ , \, \ell = 0,1, \dots, L-1 \ ,
\label{eq:vectorized}
\end{equation} 
where $U+1$ is the number of neural units in layer $\ell$, whereas $f(\cdot,\bm{w}^u_\ell,\bm{s}^u_\ell,b^u_\ell)$, for $u=0,\dots,U$, is the {\it activation} value of the corresponding neural unit $u$. Moreover, $\bm{w}^u_\ell$, $\bm{s}^u_\ell$, and $b^u_\ell$ respectively represent the weight vector, the state vector, and the bias associated with unit $u$ in layer $\ell$. It can then be seen that applying  \eq{eq:representation} to the full sequence $\bm{X}$, the representation obtained at the last layer $L-1$ represents a {\it feature matrix} $\bm{S}_{L-1}$ encoding all the temporal information contained in $\bm{X}$.\\

\noindent \textbf{Shallow Decoder:} \comment{a shallow feed forward decoder is utilized to decode the RNN output, obtaining the autoencoder depicted in Fig.~\ref{fig:rnn_encoder}. This feed forward layer decodes the embedded timeseries at the output of the RNN, and the decoding process succeeds when the output sequence matches the input RNN sequence $\bm{X}$. For this decoder, we use a linear layer with a number of units matching the size of the input vectors (i.e., $9$ samples). These units have linear activation functions and each sample of the decoded sequence is obtained as a linear combination of the RNN output.}

Formally, let $\bm{W}_D$, and $\bm{b}_D$ be the weight matrix, and the bias vector of the decoder, respectively. Then, the output sequence $\hat{\bm{X}} = [\hat{\bm{x}}_0,\hat{\bm{x}}_1,\dots,\hat{\bm{x}}_N]$, $\hat{\bm{X}} \in \mathbb{R}^{9\times {N+1}}$, is generated as shown in the following \eq{eq:decoder}:
\begin{equation}
\hat{\bm{x}}_i = 
\begin{pmatrix}
\bm{w}^0_D\cdot R_{L-1}(\bm{X}_{0,i})+b^0_D\\
\vdots \\
\bm{w}^8_D\cdot R_{L-1}(\bm{X}_{0,i})+b^8_D 
\end{pmatrix}\ , \, i = 0,1, \dots, N\ .
\label{eq:decoder}
\end{equation}
The parameter $\bm{w}^k_D$, $b^k_D$ are the weights vector and the bias of unit $k$ in the decoder, respectively. According to \eq{eq:decoder}, once the sample $\hat{\bm{x}}_N$ has been obtained, the encoder has processed the entire input sequence $\bm{X}$. If $\hat{\bm{X}} \simeq \bm{X}$, then the information ({\it features}) contained in the state matrix $\bm{S}_{L-1}$ of the output layer of the encoder nicely captures the input sequence $\bm{X}$. To improve the quality of this information, and hence that of the reconstructed sequence $\hat{\bm{X}}$, a training phase is needed, as we discuss next.\\

\noindent \textbf{Training:} To obtain encodings that are suitable for the subsequent classification task, the \mbox{RNN-Seq2Seq} module is only trained on ``normal'' walking cycles. This is because by learning to only reproduce normal gaits, when anomalous ones are given as input, the module should be unable to determine a correct embedding, and this is detected with high probability by a subsequent classifier. The objective function to minimize during the training phase is the Mean Squared Distance (MSD). Indeed, by minimizing the MSD, one maximizes the match between the input ($\bm{X}$) and the output ($\hat{\bm{X}}$) sequences. This ensures that the embedding of the input gait cycles obtained by the RNN encoder is a good higher order representation of the original timeseries.

The training is performed by means of the Back Propagation Through Time (BPTT) algorithm with no truncation~\cite{DL2016}. This is because the time span (i.e., $200$ samples in our application) of the considered timeseries does not make truncation a computational necessity. By not truncating the BPTT, we also make sure that the temporal correlation contained in the input gait cycles is fully captured, and reflected in the updates of weights and biases. To update the \mbox{RNN-Seq2Seq} weights and biases, we use a Stochastic Gradient Descent (SGD) algorithm with exponentially decaying learning rate. SGD is a common choice for training Seq2Seq models, see for example~\cite{NMT1}. To overcome the lack of flexibility of standard SGD, we introduced an exponentially decaying factor to gradually reduce the learning rate with the number of training steps. This strategy is intended to perform a fast minimization of the objective function at the beginning and to slow it down at  later iterations making these last updates more accurate. Formally, the cost function to be minimized using a gradient descent strategy is:
\begin{equation}
\begin{split}
&\mathrm{min} \displaystyle \sum_{i=0}^{N}{\sum_{j=0}^{8}{(x_{i,j}-\hat{x}_{i,j})^2}} \\
&\textrm{with respect to: }
\left\{
\begin{matrix}
\bm{W}_\ell,\ \bm{S}_\ell,\ \bm{b}_\ell & \ell = 0,\dots,L-1 \\
\bm{W}_D, \bm{b}_D & D:\ \mathrm{decoder}\\
\end{matrix}
\right.
\end{split}\ 
\label{eq:opt_encoder}
\end{equation}
\noindent where $x_{i,j}$ and $\hat{x}_{i,j}$ indicate the \mbox{$j$-th} elements of $\bm{x}_i$ and $\hat{\bm{x}}_{i}$, respectively. Index $i$ runs over the samples in a gait cycle, i.e., vector $\bm{x}_i$, and $j$ spans over the elements of this vector.

\subsection{CNN Classifier}\label{sec:class}

Next, we present the binary classification architecture that we implemented to identify anomalous gaits.

\begin{figure*}[t]
\center
\def\svgwidth{\columnwidth}
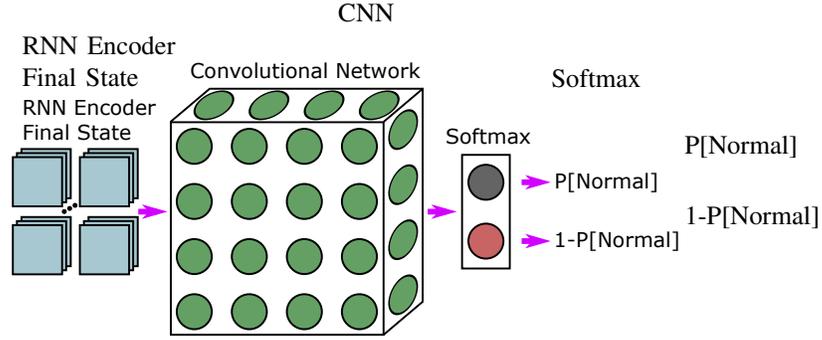
\caption{CNN-based binary classifier Architecture.}\label{fig:cnn_classifier}
\end{figure*}

Once the trained encoder has processed the whole sequence $\bm{X}$, as explained in \secref{sec:emb}, its final state $\bm{S}_{L-1}$ contains the higher order representation of the input time series. $\bm{S}_{L-1}$ is a multidimensional vector (i.e., a tensor). By suitably reshaping it, we obtain a multidimensional matrix $\tilde{\bm{S}}_{L-1} \in \mathbb{R}^{W\times Y\times Z}$, where the three dimensions $W,Y,Z$ are implementation dependent. By performing this step, on top of the temporal dependencies encoded in the original state, we determine spatial relations between the individual features of the embedding produced by the encoder. These spatial relations can be exploited by the architecture shown in \fig{fig:cnn_classifier}. The multidimensional matrix $\tilde{\bm{S}}_{L-1}$ is fed into a multilayer CNN with REctified Linear Unit (RELU) activations. Between each layer a \mbox{max-pooling} step is performed to extract the most relevant features from the kernels, and to reduce the computational complexity of the model~\cite{DL2016}. The CNN output is then flattened and fed to a \mbox{$2$-units} layer with logistic activation functions. Hence, the final classification is obtained by means of a softmax layer.

Formally, let ${\rm CNN}(\tilde{\bm{S}}_{L-1})=\bm{c} \in \mathbb{R}^M$ be the flattened output of the CNN, where $M$ is implementation dependent. Then, according to the notation used in \secref{sec:emb}, at the output of the classification layer we obtain:
\begin{equation}
\bm{s} = 
\begin{pmatrix}
s_0\\
s_1
\end{pmatrix} = 
\begin{pmatrix}
\sigma(\bm{c}\cdot \bm{w}^0_{\sigma}+b^0_{\sigma})\\
\sigma(\bm{c}\cdot \bm{w}^1_{\sigma}+b^1_{\sigma})
\end{pmatrix}\ ,
\label{eq:scores}
\end{equation}
where indices $0$ and $1$ identify the ``normal'' and ``anomalous'' gaits, respectively, and $\sigma(\cdot)$ is the sigmoid function. According to \eq{eq:scores}, $\bm{s}$ represents the scores associated with each of the two classes for a given input gait. To obtain a probability distribution of the class assignment for a given input, a softmax operation is performed as shown in \eq{eq:softmax}:
\begin{equation}
\begin{cases}
p(\bm{X} \in \textrm{normal gait} ) = \displaystyle{{\exp(s_0)}\over{\exp(s_0)+\exp(s_1)}} & \\
p(\bm{X} \in \textrm{anomalous gait}) = \displaystyle{{\exp(s_1)}\over{\exp(s_0)+\exp(s_1)}} &
\end{cases}\ ,
\label{eq:softmax}
\end{equation}
where $s_0$ and $s_1$ are the first and second component of $\bm{s}$, respectively. The result of \eq{eq:softmax} is a probability distribution and, hence, $\bm{X}$ belongs to the class with the highest probability, i.e., either ``$0$'' meaning ``normal gait'', or ``$1$'' meaning ``anomalous gait''.

To train the classifier, we used both ``normal'' and ``anomalous'' gaits that have been labeled at acquisition time. The \mbox{cross-entropy} loss applied to the softmax output was selected as the function to minimize during training. As in \secref{sec:emb}, we performed the minimization by means of Stochastic Gradient Descent (SGD). The use of RELU activations and the \mbox{cross-entropy} objective function have the advantage of mitigating the vanishing gradient issue that arises when training deep classifiers. 

\subsection{Experimental Setup and Implementation Details}\label{sec:es} 

Next, we detail the experimental setup utilized \comment{to assess the performance of our system. This setup includes:} the developed signal acquisition system (Section~\ref{sec:ALV}) to gather motion and video data from Android smartphones, the \mbox{pre-processed} dataset (Section~\ref{sec:dataset}), and the RNN/CNN based algorithms for motion learning and classification (respectively treated in Sections~\ref{sec:RNN} and~\ref{sec:CNN}). The framework is publicly available at: \href{\CodeAddress}{\CodeAddress}.\\
  
\subsubsection{Activity Logger Video Application}
\label{sec:ALV}

Inertial and video data are collected by means of a custom made Android application called ``Activity Logger and Video'' (ALV). 
ALV was tested on an Asus Zenfone $2$ featuring a $2.3$~GHz \mbox{quad-core} Intel CPU, $4$~GB of RAM and an Android $5$ ``Lollipop'' operating system. ALV requires a minimum API Level of $8$ and requests permissions for audio and video recording, camera access, Internet, and external storage. The application is used to set the acquisition parameters, collect information
about the user (age, height, gender, etc.), collect and save data into the smartphone \mbox{non-volatile} memory and, optionally, send them to a File Transfer Protocol (FTP) server. During the acquisition phase, the smartphone is carried with the \mbox{rear-facing} camera looking forward and is mounted on an \mbox{ah-hoc} made chest support as shown in \fig{fig:chest_support}(a).\\

\begin{figure}[t]
\centering
\includegraphics[scale=0.145]{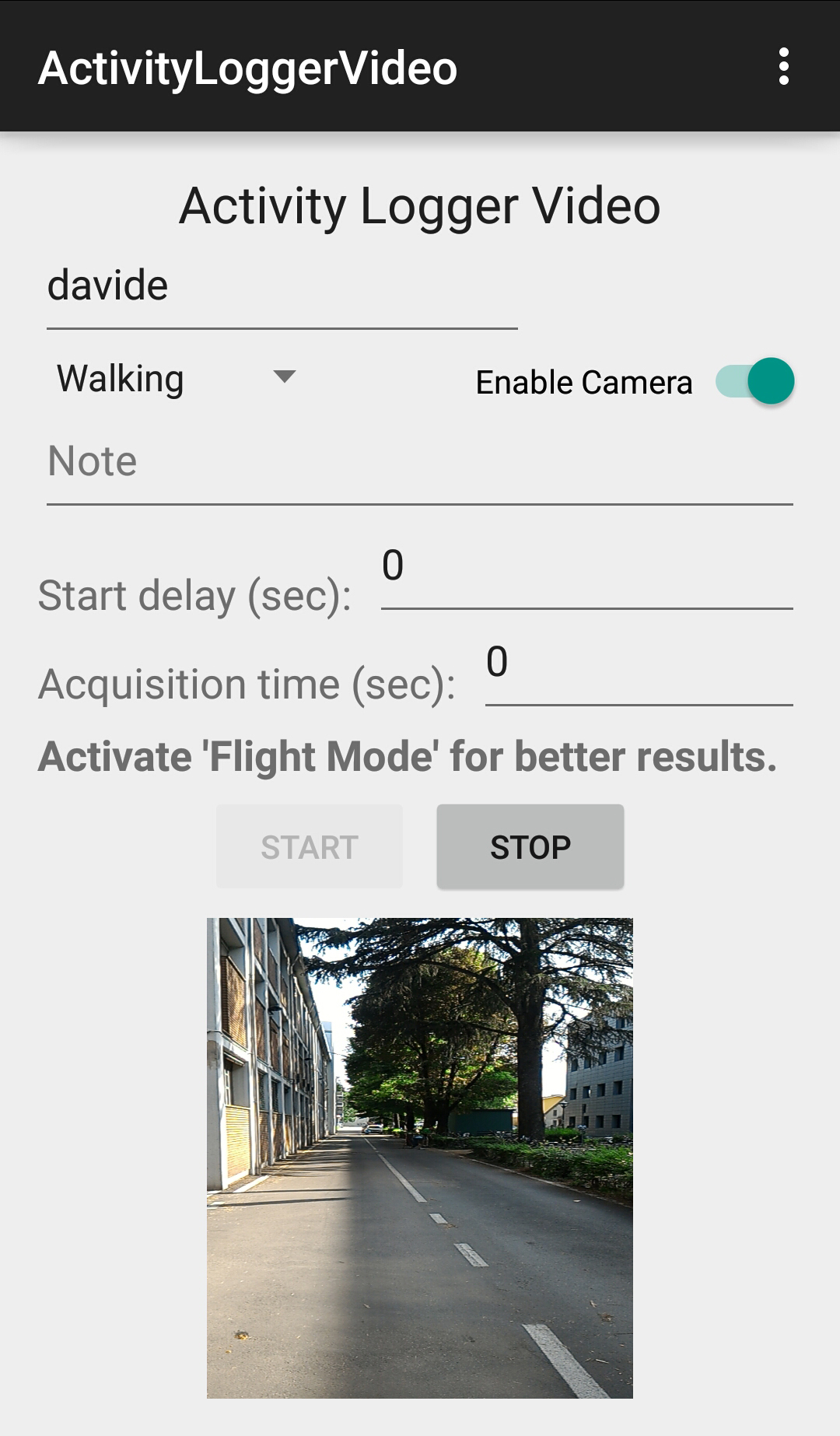}
\caption{ALV homescreen.}\label{fig:app}
\end{figure}


\subsubsection{Dataset Creation}
\label{sec:dataset}

Raw data were collected from a single healthy male subject in his twenties. We recall that our application is meant to be {\it subject-specific}, i.e., to tune itself to the gait patterns of a specific subject in order to recognize whether their walking style changes. The individual performed $141$ ``normal'' walks, and $61$ ``anomalous walks''. For the ``normal walks'' the subject walked following their natural walking habit. The corresponding patterns may contain speed variations, rotations and/or lateral oscillations, that are however to be considered normal for the purpose of this study. \comment{In fact, our objective amounts to detecting subtle variations in the walking patterns, which may be a proxy to a disease, injury, etc., that leads to a gradual or sudden degradation of motor functions.} \comment{Upon collecting the normal walks, the monitored subject was instructed to include anomalies in their walking style. This is in line with previous studies from the literature, e.g., \cite{Pogorelc-2010, Cola-2015}, where the authors emulated anomalies such as: left/right knee condition, shuffling walk (Parkinson), hemiplegia, leg pain or back pain.}

\comment{The anomalies that were considered in our present study are: shuffling walk and sliding feet, which is a typical walking attitude of a person affected by Parkinson, hemiplegic gait, which may result from a stroke~\cite{Li-2018}, unusual and subtle speed variations as the ones that arise in Alzheimer's disease. For instance, cautious gait is seen in early Alzheimer's disease: gait changes may be initially subtle and difficult to visually detect, arising with a reduction in the speed and stride of walking~\cite{Merory-2007}. Finally, we emulated anomalous trunk postures, such as those that may arise in the diabetic gait~\cite{Sawacha-2009}.} \comment{Our purpose in the present study is to propose a very accurate detection engine, assessing whether gait anomalies of some sort are present in the walking patterns. Their further classification into the corresponding anomaly type is not considered. We remark that this task can be accomplished by replacing the binary CNN classifier with a CNN (or any other classification scheme) that uses the RNN state to discriminate among the different anomaly classes. This further assessment would require the use of clinical data and to focus on one or on several specific pathologies, which is left as a future endeavor.}  

For each walk, accelerometric, gyroscopic, and video data were collected and \mbox{time-synchronized}. The collected data and timestamps were then processed, segmented, and normalized as described in \secref{sec:dp}, resulting in $7,941$ \comment{``normal gait cycles''}, and $2,744$ \comment{``anomalous gait cycles''}. Each gait is represented through a $9\times 200$ real matrix. From the ``normal'' \comment{gait cycles}, $4,966$ have been used for training and testing the RNN-Seq2Seq model discussed in \secref{sec:emb}. The remaining ``normal'' \comment{gait cycles} and all the ``anomalous'' ones have been used for training and testing the classifiers.\\ 

\subsubsection{RNN-Seq2Seq model implementation and training}
\label{sec:RNN} 
The RNN-Seq2Seq model classifier has been implemented in \mbox{Python-$3.6$}, using the \texttt{TensorFlow} library. The encoder is made of $2$ bidirectional LSTM layers. Each layer has $512$ LSTM cells with forget gates and peepholes connections. To implement this model, we used the \texttt{tf.contrib.rnn.LSTMBlockCell} class with the \texttt{use\_peepholes} flag set to \texttt{True}. The decoder is made of a single dense layer with $9$ units with linear activation functions. It has been implemented by means of the \texttt{tf.layers.Dense} class with the \texttt{activation} parameter set to \texttt{None}. For training purposes, the RNN encoder has been wrapped using a \texttt{tf.contrib.rnn.DropoutWrapper} wrapper with output keep probability of $0.8$. The objective function has been defined by means of the \texttt{tf.losses.mean\_squared\_error} loss, and the training phase lasted for $21$ epochs. The learning rate has been halved every $1,000$ training steps on \mbox{mini-batches} of size $16$ samples. Moreover, we implemented a gradient clipping step, to prevent the exploding gradient issue. The gradients have been clipped with respect to a maximum norm of $5$. The training was executed on a desktop PC equipped with a NVidia \mbox{Titan-X} GPU. The training dataset was made by $90\%$ of the ``normal'' \comment{gait cycles}, the remaining $10\%$ has been used for testing purposes. It is worth noting that these data have been used only to train the RNN-Seq2Seq model, and have not been used for the subsequent training and testing of the classifier. This is to prevent the encoder from generating too good representation of the ``normal'' \comment{gait cycles}, hence biasing the following classification step.\\

\subsubsection{CNN classifier implementation and training}
\label{sec:CNN}
The CNN classifier has been implemented in Python-$3.6$, utilizing the \texttt{TensorFlow} library. We implemented one convolutional layer with the \texttt{tf.nn.conv2d} module. For this we used \texttt{[10, 6, 8, 16]} shaped filters initialized with:\\ \texttt{tf.truncated\_normal\_initializer} with standard deviation \texttt{stddev = $0.05$}. The output of the convolutional layer is then processed by a max pooling layer implemented through \texttt{tf.nn.max\_pool}. For this, we used kernels of size \texttt{[1, 4, 4, 1]}, and strides of length \texttt{[1, 2, 2, 1]}. The resulting tensor is then flattened and processed by a \mbox{$2$-units} \texttt{tf.layers.dense} later with activations set to \texttt{None} (i.e., with linear activations). This is because, for computing the \mbox{cross-entropy} loss we used the \texttt{tf.nn.softmax\_cross\_entropy\_with\_logits} function that applies the logistic function directly to the output of a linear layer. The loss has been minimized using SGD with exponentially decaying learning rate. To this end, the learning rate has been halved every $1,000$ training steps. The training has been performed for $11$ epochs with mini-batches of size $16$. The classification dataset consists in mixed ``normal'' and ``anomalous'' \comment{gait cycles}, with a ratio of about $1:1$. The training process involved a first encoding phase by means of the \mbox{pre-trained} RNN-Seq2Seq model. The resulting final state is then reshaped as described in \secref{sec:class}, and fed to the classifier. For the training, we used $90\%$ of the previously shuffled dataset. The remaining $10\%$ has been used for testing.\\

\subsubsection{SVM classifier implementation and training}
The SVM classifier has been implemented in Python-$3.6$ with the support of the \texttt{ScikitLearn} library. We chose a Gaussian RBF kernel, and we fed the model with flattened gait cycles. The model has been trained on the same dataset used for the CNN classifier. The training set was composed of $90\%$ of the data. The trained model has then been tested on the remaining $10\%$ of the data. It is worth noting that, usually, SVM models require less training data with respect to \mbox{NN-based} ones. Hence, by using the same \mbox{train-test} split ratio for all the architectures, we are actually slightly advantaging the SVM classifier.

\comment{We remark that the two classifiers (i.e., the CNN-based one and the SVM one) have been compared against the same previously unseen test set. The considered test set is composed of $572$ gait cycles split in the following way:
\begin{itemize}
	\item $49.3\%$ of normal gait cycles;
	\item  $50.7\%$ of anomalous gait cycles.\\
\end{itemize}}

\subsubsection{Synopsis of the implementation parameters}
Here, we recap the main parameters of the implemented models.\\

\myparstart{RNN-Seq2Seq model}
\begin{table}[H]
\begin{center}
\begin{tabular}{|c|c|}
\hline
\textbf{Training Data} &  $4,469$ ``normal'' \comment{gait cycles}\\
\hline
\textbf{Test Data} & $497$ ``normal'' \comment{gait cycles}\\
\hline
\textbf{Recurrent Layers} & $2$, bidirectional \\
\hline
\textbf{Units Per Layer} & $512$ LSTM cells \\
\hline
\textbf{Initial Learning Rate} & $0.01$ \\
\hline
\textbf{Decay Steps} & $1,000$ \\
\hline
\textbf{Training Epochs} & $21$ \\
\hline
\textbf{Max Gradient Norm} & $5$ \\
\hline
\end{tabular}
\caption{RNN-Seq2Seq model (\secref{sec:emb}) Implementation Parameters}
\end{center}
\end{table}


\myparstart{CNN-based classifier}
\begin{table}[H]
\begin{center}
\begin{tabular}{|c|c|}
\hline
\textbf{Training Data} &  $5,147$ mixed \comment{gait cycles}\\
\hline
\textbf{Test Data} & $572$ mixed \comment{gait cycles}\\
\hline
\textbf{Convolutional Layers} & $1$ \\
\hline
\textbf{Convolutional Filters} & [$10,\ 6,\ 8,\ 16$] \\
\hline
\textbf{Convolutional Strides} & [$1,\ 2,\ 2,\ 1$] \\
\hline
\textbf{Max Pooling Layers} & $1$ \\
\hline
\textbf{Max Pooling Kernel} & [$1,\ 4,\ 4,\ 1$] \\
\hline
\textbf{Max Pooling Strides} & [$1,\ 2,\ 2,\ 1$] \\
\hline
\textbf{Initial Learning Rate} & $0.01$ \\
\hline
\textbf{Decay Steps} & $1,000$ \\
\hline
\textbf{Training Epochs} & $11$ \\
\hline
\end{tabular}
\caption{Convolutional Classifier (\secref{sec:class}) Implementation Parameters}
\end{center}
\end{table}

\myparstart{SVM-based classifier}
\begin{table}[H]
\begin{center}
\begin{tabular}{|c|c|}
\hline
\textbf{Training Data} &  $5,147$ mixed \comment{gait cycles}\\
\hline
\textbf{Test Data} & $572$ mixed \comment{gait cycles}\\
\hline
\textbf{SVM Kernel} & Gaussian RBF\\
\hline
\end{tabular}
\caption{SVM Classifier (\secref{sec:class}) Implementation Parameters}
\end{center}
\end{table}
\section{Results}
\label{sec:res}

\begin{table}[tbp!]
\begin{center}
\begin{tabular}{|c|c|c|}
\hline
& \textbf{\comment{predicted} ``normal''} & \textbf{\comment{predicted} ``anomalous''}\\
\hline
\textbf{``normal''}  & $47.38\%$ & $1.92\%$\\
\hline
\textbf{``anomalous''}  & $0.00\%$ & $50.70\%$\\
\hline
\end{tabular}
\caption{SVM Classifier: Confusion Matrix on the Test Set.}\label{tab:cm}
\end{center}
\end{table}

To assess the classification performance of the proposed method, we compared it to that of the SVM classifier of \secref{sec:es}. The SVM model achieved a classification accuracy of $98.077\%$ on the test set. The corresponding confusion matrix is shown in \tab{tab:cm}, from which two main conclusions can be drawn. First, even a standard model achieves a high classification accuracy. Second, the SVM provides ``conservative'' results. This means that it is more likely that a ``normal'' gait is classified as anomalous than the opposite. This is also a desirable feature for healthcare monitoring applications, because no potentially dangerous conditions are misclassified.

The classification accuracy on the test set that we obtained with the proposed \mbox{2-step} model (RNN as feature extractor followed by a CNN classifier) is $\bm{100\%}$. This improvement, together with the ability of neural networks to perform online learning (i.e., updating the model parameters while being used), make the proposed architecture a viable choice for gait anomaly detection in free living conditions. Moreover, it is worth noting that, once the initial training of the RNN-Seq2Seq model and the CNN-based classifier has been completed, the resulting model can be run on off-the-shelf smartphones as the Asus Zenfone $2$ used in this work.\\

\paragraph{Discussion} We found the problem of detecting anomalies in gait cycles to be a difficult one. Although here we only describe the final design, which worked satisfactorily in all our experiments, prior to this we tried several other ways of combining neural networks and classifiers. For example, we experimented with a system architecture where the RNN is trained to predict the next sample $\bm{x}_i \in \mathbb{R}^9$, given the (observed) previous samples in the current cycle $\bm{x}_0, \bm{x}_1, \dots, \bm{x}_{i-1}$. We thus evaluated the prediction error \mbox{$\bm{e}_i = \bm{x}_i - \hat{\bm{x}}_i$} for each sample $i$, with $i=0,1,\dots,N+1$. Hence, for each gait cycle, we evaluated the Mean Square Error (MSE) and its standard deviation for each element of $\bm{e}_i$, leading to a {\it cycle descriptor} of size $2 \times 9$. This cycle descriptor was then used to discriminate between normal and anomalous cycles using an SVM classifier. In the test phase, this technique has led to a percentage of normal cycles that were misclassified as anomalous of $5.8$\%, and to a percentage of anomalous cycles that were misclassified as normal of $2.9$\%. The percentage of correctly classified cycles were $42.7$\% and $48.6$\% for normal and anomalous walks, respectively. 


\section{Conclusions}\label{sec:conc}

In this work, we have explored new methods for the automated classification of gait cycles from multimodal motion data consisting of inertial (accelerometer and gyroscope) and video signals. The proposed solution is {\it subject-specific}, as we purposely learn the way in which a subject walks, with the goal of correctly classifying normal gaits from anomalous ones, where ``anomalous'' mean containing subsequences that are usually not present in the normal walks. To the best of our knowledge, this is the first work that systematically addresses this problem through the fusion of inertial and video data. Also, we utilize advanced deep neural network models, namely, Recurrent Neural Networks (RNN), to capture the statistics underpinning the motion data, through an architecture that combines recurrent and convolutional designs. The system that we put forward is lightweight, as data can be conveniently acquired from a smartphone device with no user configuration required. The feature extraction block is trained in an unsupervised manner (thanks to RNN) and labeled data is only needed for the training of the final classifier. \comment{Our results reveal that the final architecture is very effective, being able to correctly capture the correlations in the walking patterns of the monitored subject and to detect all the anomalous gaits. Our work can be extended in several ways. For instance, we may assess which type of anomaly affects the data, or use a similar design for activity recognition tasks. Also, this trajectory learning approach can be very useful to quantify the progress of a disease affecting motor functions or the benefits of rehabilitation therapies.}

\section*{Acknowledgment}
This work has been supported by the University of Padova through the project CPDA 151221 ``IoT-SURF''. Any opinions, findings, and conclusions herein are those of the authors and do not necessarily represent those of the funding institution.

\section*{Ethics Statement} 
\comment{The present study does not involve the use of clinical data. The learning and classification algorithms have been trained using walking patterns from a healthy subject, who emulated walking anomalies as explained in Section~\ref{sec:es}. The subject has provided a written consent to use their walking data for the numerical assessment that we carried out in Section~\ref{sec:res}.}     

\bibliographystyle{IEEEtran}
\bibliography{bibliography}



\end{document}